Ya.Ye. Khaustov[1], D. Ye. Khaustov[1], Ye.Ryzhov[1], E. Lychkovskyy[2], Yu. A. Nastishin[1]

[1]*Hetman Petro Sahaidachnyi National Army Academy*
[2]*Danylo Halytsky Lviv National Medical University*

# FUSION OF VISIBLE AND INFRARED IMAGES VIA COMPLEX FUNCTION

*We propose an algorithm for the fusion of partial images collected from the visual and infrared cameras such that the visual and infrared images are the real and imaginary parts of a complex function. The proposed image fusion algorithm of the complex function is a generalization for the algorithm of conventional image addition in the same way as the addition of complex numbers is the generalization for the addition of real numbers. The proposed algorithm of the complex function is simple in use and non-demanding in computer power. The complex form of the fused image opens a possibility to form the fused image either as the amplitude image or as a phase image, which in turn can be in several forms. We show theoretically that the local contrast of the fused phase images is higher than those of the partial images as well as in comparison with the images obtained by the algorithm of the simple or weighted addition. Experimental image quality assessment of the fused phase images performed using the histograms, entropy shows the higher quality of the phase images in comparison with those of the input partial images as well as those obtained with different fusion methods reported in the literature.*

## Introduction

The fusion of images is the synthesis of an image by a combination of data from several partial images, which carry different information about the same scene. Partial images can be obtained either by the same camera (most often visual CCD-camera) collecting images at different conditions (such as focusing, direction of observation, time moment) or by different detectors registering signals of different nature (electromagnetic or ultrasound waves in transmitting or reflecting modes, distribution of temperature, pressure or other physical parameters) or signals of the same nature, most often electromagnetic waves in different regions of electromagnetic spectrum: γ-rays, X-rays, UV, visible (Vis), infrared (IR) light, radio-waves. Up to date, there is no universal material, which can be effectively used as a sensor of signals in the whole electromagnetic spectrum. For different spectral regions, one uses detectors, working on different physical principles. Consequently, the fusion of analogous images from different spectral regions is hardly achievable, especially taking into account that until recently analogous pictures were obtained in the form of a paper or film photograph, obtained by the chemical development process or as an image on a scintillation monitor. Although up to date still there is no universal detector capable for covering whole electromagnetic spectrum, remarkably, nowadays for all the electromagnetic regions there are detectors designed in a form of pixel matrix of sensing elements such that in response to the irradiation by the electromagnetic wave each pixel produces own electrical signal.

Therefore, an image can be electronically saved in the form of a computer file. Digitalization of pixel electrical signals drastically changed the situation in imaging, becoming a basis for different operations with images that now are covered by the term the image processing. In digital pixel format an image, called the electronic image, is presented by a pixel table of brightness values for the detectors working in different spectral regions. The importance of the electronic format of imaging is hard to overestimate. The fusion of images is one among many other possibilities in image processing, which became possible and handy due to the electronic format of imaging. Importantly, due to the electronic format, the image processing can be done "à la source", i.e. within the same device immediately after recording the raw data from the detector. The latter is important regarding the development of compact cameras capable of collecting partial images in different spectral regions and displaying on the monitor a single fused image of improved quality in comparison with the raw partial images. Such compact smart cameras are in great demand in situations when bulky registration systems cannot be used. Corresponding examples are many: for medicine, when monitoring the state of a patient out of a hospital or of a non-transportable patient, using different detectors (X-ray, magnetic resonance tomography, ultrasound, etc.); for military needs, for example in a target sightseeing systems of motor vehicles (tanks, aircrafts, ships, submarines) employing visual, thermal and radar imaging; for space surveillance; geodesic and marine reconnoiter; etc.

## Analysis of recent publications

Once the partial images are fused, one has to assess the quality of the fused image. Several methods have been proposed for the estimation of the quality of a fused image (EQFI). These methods can be divided into two groups, namely: subjective and objective methods [1]. In this paper, we are interested only in the objective methods providing quantitative indices for EQFI. Our analysis of the literature proposing such EQFI indices [2] shows that as a rule they are based on the notion of the integral contrast, which is calculated for the area of the image, much larger than the area of a single pixel. Recently we have argued that there are situations at which the notion of the local contrast is more appropriate for the adequate EQFI result. The notion of local contrast is more appropriate when the partial images contain large areas of zero or very low brightness. Such a situation is typical for visual images taken at low lightening, frequently for thermal images and practically always for radar images. It is clear that in such cases the value of the integral contrast would be greatly affected by the irrelevant information from the areas of low brightness and, thus, the integral contrast should be replaced by the local contrast, which is calculated for a single pixel with respect to the pixels of its nearest vicinity. For this reason, we will use the approach of local contrast in the EQFI procedure (see [2] for details).

The most obvious and most commonly used algorithm for the fusion of two (or more) images is the addition of their electronic tables. Many other sophisticated image fusion algorithms have been developed [2]. Some of them have been designed specifically for the fusion of visible and infrared images [3-6]. An extended list of references on the infrared and visible image fusion methods and their applications can be found in the recent review by Jiayi Ma, Yong Ma and Chang Li [7], where eighteen fusion methods are compared using nine assessment metrics. All of them have own advantages, limitations, and drawbacks and, thus, the development of new fusion algorithms that provide high quality fused images is still in demand.

Most of the available image fusion algorithms by default have been designed for the office work, requiring powerful computers and a specially trained operator with solid knowledge in programming. In addition, since most of them were developed for civil needs implying operations involving elements of art, personal characteristics of the operator (such as the professional art knowledge, intuition, and ability for decision-making, personal skills in programming) play an important role in the final appearance of the fused image. Only a few such algorithms can be classified as objective algorithms, which are free of the subjective influence of the operator, and are simple in use and non-demanding in terms of the computer capabilities. The algorithms of simple and weighted fusion fulfill these requirements.

However, recently we have shown [2] that the simple and weighted fusion algorithms have at least two significant drawbacks. The first drawback is that the local contrast of an image synthesized using the simple and weighted fusion algorithms are always lower than the contrast of one of the partial images. Moreover, we have demonstrated that the simple and weighted fusions are not always reasonable. In some cases, the image fusion worsens the image in comparison with that of original partial images. Our analytical relations for the local contrast of the fused image as a function of the local contrasts of partial images indicate [2] that the local contrast of a pixel in the fused image can vanish when fusing two partial images from the visible and thermal cameras. Vanishing of the local contrast of the pixel in the fused image is due to the fact that the local contrasts on the visible and thermal images are of opposite signs. As a result, an object which is visible on both partial images can become invisible on the fused image. This is the second drawback of the simple and weighted fusion algorithms.

## Aim of the article

In this paper, we propose the image fusion algorithm of complex function, which is free of both these drawbacks, at the same time remaining to be simple in use and non-demanding concerning the power of the computer. The proposed image fusion algorithm of the complex function is a generalization for the algorithm of conventional algorithms of image addition in the same way as the addition of complex numbers is the generalization for the addition of real numbers.

## Results

**1. Description of the image fusion algorithm of complex function**

It is important to recall, that the digital images are defined by their pixel brightness tables. Before fusing the partial images, one has to perform alignment (also called registration) of their brightness tables. The image alignment procedure consists in the formatting of the pixel brightness tables of the partial images such that the pixels corresponding to the same elements of the partial images (key points) become of the same coordinates in their brightness tables [8,9]. In this and the following sections, we assume that the process of alignment of partial images is successful and the partial images from the channels $A$ and $B$ are ready for fusion. Let the cells of the pixel brightness table be denoted by natural numbers $x$ and $y$ in the row and column of the table, respectively. To increase the efficiency of the fusion of images specified by their brightness tables $u^A(x,y)$ and $\upsilon^B(x,y)$ obtained from two channels $A$ and $B$, we propose to present the fused image $\psi(x,y)$ in the form of a complex function built of the brightness tables $u^A(x,y)$ and $\upsilon^B(x,y)$, one of which will be chosen as the real and the other one as the imaginary part of the function $\psi$. There are two possibilities for this, namely:

$$\psi_{neg}(x,y) = u(x,y) + i\upsilon(x,y), \quad (1)$$
$$\psi_{pos}(x,y) = \upsilon(x,y) + iu(x,y), \quad (2)$$

where $i$ is the imaginary unit such that $i^2 = -1$, $u(x,y) = w^A(x,y)u^A(x,y)$, $\upsilon(x,y) = w^B(x,y)\upsilon^B(x,y)$ are the weighted values of brightness, according to their weight coefficients, $w^A$ and $w^B$, such that $\sqrt{(w^A)^2 + (w^B)^2} = 1$. Therefore, by definition, the proposed complex image fusion algorithm is a generalized prototype of the weighted pixel averaging algorithm [1], but, as we show below, with much wider possibilities for the image fusion. The weighting factors $w^A$ and $w^B$ determine the degree of importance (relevance) of the information from the channels. The values of the weight coefficients can be entered either manually, as decided by the operator, or the optimal values of the weight coefficients can be determined, for example, by the calculation of their weight with respect to the averaged brightness values of the corresponding pixels of the partial images, or employing special methods, such as, for example, principal component analysis (PCA) [10]. The meaning of the subscript indices *neg* and *pos* in Eq. (1) and (2) which correspond to the negative and positive images will be identified later in this section.

One of the advantages of the proposed algorithm of complex function is in the wider possibilities for the combination of the information carried by the partial images in comparison with the simple and weighted fusion. Indeed, the complex forms of Eq. (1), (2) of the fused image implies that in the case of several $l$, and $m$ images, respectively from the two channels, the fused image is also a complex function such that the images from different channels are sorted and added separately and then the two resulting images from the two channels are fused by the algorithm of complex function, namely:

$$\Psi_{neg}(x,y) = \sum_{j=1}^{l} w_j^A u_j(x,y) + i\sum_{k=1}^{m} w_k^B \upsilon_k(x,y),$$
$$\Psi_{pos}(x,y) = \sum_{k=1}^{m} w_j^B \upsilon_j(x,y) + i\sum_{j=1}^{l} w_j^A u(x,y). \quad (3)$$

Different images from the same channel can be obtained with a different focus, exposure (shutter speed), or in dynamic mode for different delay times between images in the case of moving objects. The complex function $\Psi_{neg,pos}(x,y)$ can be represented in the exponential form [11]

$$\Psi_{neg,pos}(x,y) = |\Psi_{neg,pos}(x,y)| e^{i\varphi_{neg,pos}(x,y)}, \quad (4)$$

or in the trigonometric form

$$\Psi_{neg,pos}(x,y) = |\Psi_{neg,pos}(x,y)|\{\cos[\varphi_{neg,pos}(x,y)] + i\sin[\varphi_{neg,pos}(x,y)]\}, \quad (5)$$

where

$$|\Psi_{neg,pos}(x,y)| = \left\{\begin{array}{l}\left(\operatorname{Re}[\Psi_{neg,pos}(x,y)]\right)^2 + \\ +\left(\operatorname{Im}[\Psi_{neg,pos}(x,y)]\right)^2\end{array}\right\}^{\frac{1}{2}} \quad (6)$$

is nothing else but the brightness of the complex signal of the pixel with the coordinates $(x,y)$, and $\varphi_{neg,pos}(x,y)$ is the phase of the complex image such that

$$\varphi_{neg,pos}(x,y) = \arctan\left[\frac{\operatorname{Im}\{\Psi_{neg,pos}(x,y)\}}{\operatorname{Re}\{\Psi_{neg,pos}(x,y)\}}\right]. \quad (5)$$

Substitution of Eqs. (3) in Eqs. (7) gives

$$\varphi_{neg} = \arctan\left[\frac{\sum_{k=1}^{m} w_k^B \upsilon_k(x,y)}{\sum_{j=1}^{l} w_j^A u_j(x,y)}\right],$$
$$\varphi_{pos} = \arctan\left[\frac{\sum_{j=1}^{l} w_j^A u_j(x,y)}{\sum_{k=1}^{m} w_k^B \upsilon_k(x,y)}\right]. \quad (6)$$

It follows from Eqs. (7) that the phase $\varphi_{neg,pos}(x,y)$ takes the values between 0 and $\pi/2$. The exponential and trigonometric forms, of Eq. (4)-(7) imply that a new fused image can be built by the two independent algorithms or as a combination of both. Namely: in one case, the new image is obtained as the amplitude $|\Psi_{neg,pos}(x,y)|$ map, and in the other one as a phase $\varphi_{neg,pos}(x,y)$ map. One can also build a new image via several combinations of parameters $|\Psi_{neg,pos}(x,y)|$ and $\varphi_{neg,pos}(x,y)$, for example:

$$2\operatorname{Re}\{\Psi_{neg,pos}(x,y)\}\operatorname{Im}\{\Psi_{neg,pos}(x,y)\}$$
$$= |\Psi_{neg,pos}(x,y)|^2 \sin 2\varphi_{neg,pos}(x,y); \quad (7)$$

$$\left(\operatorname{Re}\{\Psi_{neg,pos}(x,y)\}\right)^2 - \left(\operatorname{Im}\{\Psi_{neg,pos}(x,y)\}\right)^2$$
$$= |\Psi_{neg,pos}(x,y)|^2 \cos 2\varphi_{neg,pos}(x,y). \quad (8)$$

It is worth noting that the combinations $\operatorname{Re}\{\Psi_{neg,pos}(x,y)\} = |\Psi_{neg,pos}(x,y)|\cos\varphi_{neg,pos}(x,y) \quad (9)$

$$\text{Im}\{\Psi_{neg,pos}(x,y)\} = |\Psi_{neg,pos}(x,y)|\sin\varphi_{neg,pos}(x,y) \quad (10)$$

do not bring new images, since they result in nothing else but in the partial images $u(x,y)$ or $\upsilon(x,y)$, respectively. The effectiveness of the above mentioned and other possible algorithms can be determined via the calculation of local contrast of the pixels in these images.

By definition [2], the local contrast $k^A$ (or $k^B$) of the pixel with the coordinates $(x,y)$ with the brightness $u(x,y)$ in the image from the channel $A$ (or, respectively, with the brightness $\upsilon(x,y)$ of the image from the channel $B$) with respect to its close neighbor pixel with the coordinates $(x+\Delta x, y+\Delta y)$ is of the form

$$k^A = \frac{\Delta u}{\bar{u}} \text{ (or } k_d^B = \frac{\Delta \upsilon}{\bar{\upsilon}}) \quad (11)$$

where

$$\begin{aligned}
\Delta u &= u(x+\Delta x, y+\Delta y) - u(x,y); \\
\bar{u} &= \frac{u(x+\Delta x, y+\Delta y) + u(x,y)}{2}; \\
\Delta \upsilon &= \upsilon(x+\Delta x, y+\Delta y) - \upsilon(x,y); \\
\bar{\upsilon} &= \frac{\upsilon(x+\Delta x, y+\Delta y) + \upsilon(x,y)}{2}.
\end{aligned} \quad (12)$$

Below we evaluate the local contrast of the phase images and show that the algorithms based on the phase component, Eq. (7) provide higher contrast in comparison with those of the simple and weighted fusion, at which the brightness values of the corresponding points are added, respectively, as $\psi^s = u + \upsilon$ and $\psi^w = w^A u + w^B \upsilon$ with the corresponding weighting coefficients $w^A$ and $w^B$. The contrast of the amplitude images $|\Psi_{neg,pos}(x,y)|$, given by Eq. 6, will be evaluated elsewhere.

**2. Contrast of images fused by the algorithm of complex function**

As it was already mentioned above, the fusion of images by the algorithms of complex function (1), (2) provides the possibility of constructing a fused image in, at least, two forms: amplitude (6) and phase (7). For two channels (visible and thermal), the view of the amplitude ($a$-) image does not depend on the choice, which of the two partial images (visible or IR) is the real part and which is the imaginary part of the fused image $\psi$, see Eq.(6). For the phase images, depending on the choice which of the images (visual or IR) is the real part and which is the imaginary part of the fused complex image $\psi$, we obtain either the negative, Eq.(1), or the positive, Eq. (2), image. In turn, the phase images can be built either in the form of a tangent of the phase ($t$-image):

$$t_{neg} = \tan\varphi_{neg} = \frac{\upsilon}{u}, \quad (13)$$

$$t_{pos} = \tan\varphi_{pos} = \frac{u}{\upsilon} = t_{neg}^{-1}, \quad (14)$$

or in the form of a phase argument ($\varphi$-image):

$$\varphi_{neg} = \arctan\left(\frac{\upsilon}{u}\right), \quad (15)$$

$$\varphi_{pos} = \arctan\left(\frac{u}{\upsilon}\right). \quad (16)$$

It is worth noting that it follows from (13), (14) and (15), (16) that $\varphi_{pos} = (\pi/2) - \varphi_{neg}$. The subscript index *neg* in equations (13) and (17) corresponds to the fusion of images according to the algorithm (1). Namely, by selecting a signal from the visible channel as the real part, and the IR signal as the imaginary part of the complex image, we obtain the images $t_{neg}$ and $\varphi_{neg}$. Respectively, the subscript index *pos* in Eqs. (14), (16) corresponds to the fusion by Eq. (2). Namely, by choosing the IR signal $\upsilon$ as the real part and the signal $u$ from the visible channel as the imaginary part, we obtain the images $t_{pos}$ and $\varphi_{pos}$. From expressions (13), (14) we find that the contrast of the *t*-image, which is given by the ratio of the signals, Eq.(13), is of the form

$$k_{neg}^t = \frac{k^B - k^A}{1 - \frac{1}{4}k^A k^B} \quad (17)$$

It is worth noting that as a rule the image fusion is needed when the brightness and, consequently, the contrast of the partial images are low. At good enough visibility, thereby at high enough contrast of the image from the visible channel, there is no need to use an image from the IR channel, and vice versa. In conditions of poor visibility in the visible channel, there is no reason to expect that the contrast of the IR image will be much higher than the contrast of the visible image. For such a case, even for relatively high contrast values, let's say $k^A = -k^B = 2/3$ one has $k^A k^B/4 = 1/9 \ll 1$. Therefore, for typical images for which the fusion is needed (i.e. at $(k^A k^B/4) \ll 1$), the expression (17) takes the form:

$$k_{neg}^t \approx k^B - k^A \quad (18)$$

Since the contrasts of images from the visible and infrared channels are of opposite signs such that $k^A > 0$, and $k^B < 0$, then it follows from (18) that in the phase image the absolute values of the contrasts of the visible and infrared images are added and, the sign of the contrast of the phase image is negative, coinciding with the sign of the IR image, which stands in the numerator of the ratio (13), and which determines the positive or negative appearance of the phase image. In this paper, we adopt a convention according to which the term "negative image" corresponds to the negative contrast of the key points. Respectively, the positive contrast corresponds to the positive image. For the very this reason, the *negative t-image* and the corresponding variables, starting with Eq. (1) and further, throughout the text, are denoted by the subscript *neg*.

Alternatively, if, the fused image is built by the algorithm (2) such that the image from the IR channel is the real part and the image from the visible channel is the imaginary part of the fused complex image, then the contrast of the $t_{pos}$ – image built by the algorithm (14) is of the form

$$k^t_{pos} = -k^t_{neg} \approx k^A - k^B \qquad (19)$$

Taking into account that $k^A > 0$, and $k^B < 0$, we conclude that, at the phase fusion by the algorithm (2), (14) the absolute values of the contrasts $k^A$ and $k^B$ of the partial images fused according to the equation (19) are added. The sign of the contrast of the fused complex image will be positive. For this reason, we call the fused complex image obtained by the algorithm (14), a *positive t–image*. For the very this reason, the corresponding variables, starting with Eq. (2) and further, throughout the text, are denoted by the subscript *pos*.

Therefore, a general conclusion is that the contrast of both positive and negative $t$–images is always higher than the contrast of the partial images, whereas, according to [2], at the simple and weighted fusion, the local contrast of the fused image is always lower than the contrast of one of the two partial images. In particular, the contrast of the $t$–images fused by the algorithms (13), (14), is doubled by its absolute value when $k^A = -k^B$, whereas at the simple addition of such partial images the contrast of the complex image becomes zero, which means that the target pixel point becomes invisible in the fused image.

Contrasts $k^{\varphi}_{neg}$ and $k^{\varphi}_{pos}$ of the $\varphi-$images are complicated functions of the contrasts of the partial images as well as of the values of their brightness. Our numerical calculations show that the contrast of the $\varphi-$images is lower than the contrast of the *t*-image, but higher than the contrasts of the partial images. Fig.1 illustrates these statements for the two model partial images of the same target in the form of a small square on the background of a larger square. In the first partial image (on left in Fig. 1a), the small target square of brightness $u_1 = 0.4$ is visible on the background of the larger brighter square of brightness $u_2 = 0.5$. From Eqs. (13), (14) one finds that the local contrast of the smaller square is $k^A = 0.2(2)$. It should be noted that, as a rule, an object of interest (a target) is seen on the image obtained in the Vis light, as a dark object on a bright background. The latter implies the positive contrast of the target and, thus, a dark small square on the background of the brighter larger square (Fig.1a on left) of the contrast $k^A = 0.2(2)$, models an image from the visible channel *A*. In the second partial image (in the middle of Fig. 1a), the small brighter target square of brightness $\upsilon_1 = 0.5$ is seen on the background of the larger but darker square of brightness $\upsilon_2 = 0.4$. Substitution of these brightness values in Eqs. (13), (14) gives $k^B = -0.2(2)$. As a rule, on an image from the thermal camera (IR channel) a target is seen as a bright target on a dark background, and, thus, a brighter small square (the target) on the background of the darker larger square (in the middle of Fig.1a) of the contrast $k^B = -0.2$ (2), models an image of the same target from the IR channel *B*.

First, we fuse these two partial images by the algorithm of simple addition (also called simple fusion) and obtain a new image:

$$\psi^S = u + \upsilon. \qquad (20)$$

In [2] we have shown that for the image $\psi^S$ obtained according to the simple fusion algorithm, Eq.(22), the local contrast of the target is

$$k^S = \omega^u k^A + \omega^\upsilon k^B, \qquad (21)$$

where $\omega^u = \overline{u}/(\overline{u}+\overline{\upsilon})$ and $\omega^\upsilon = \overline{\upsilon}/(\overline{u}+\overline{\upsilon})$ are the weights of the $u-$ and $\upsilon-$images respectively. For this pair of images (on left and in the middle in Fig. 1a) we have $\omega^u = \omega^\upsilon = 0.5$, $k^A = -k^B = 0.2(2)$, and therefore, by simple fusion according to Eq.(22), from Eq. (21) we obtain a zero contrast $k^s = 0$ of the target, which becomes invisible in the fused image (on right in Fig. 1a).

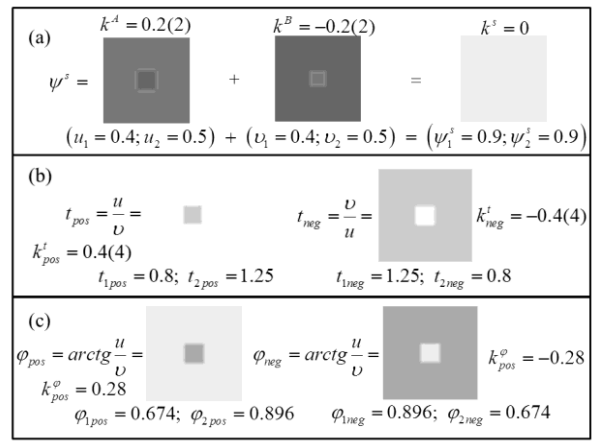

*Fig. 1.* **Comparison of the model images obtained by the simple addition algorithm (a) to the positive and negative phase *t*- images (b) and $\varphi$-images (c)**

The phase $\varphi$- and $t$-images do not have such a drawback. Indeed, the contrasts of the $t$-images obtained by the fusion of the same partial images from the visible and infrared channels according to the algorithms given by Eq. (13), shown on left in Fig. 1b, and by Eq. (14), shown on right in Fig. 1b, are defined by Eqs. (18), (19), respectively. Thus the absolute values of the local contrast for both the positive and negative $t$-images are the sum of the absolute values of the contrasts of the partial images. Therefore, the contrasts of the fused positive and negative $t$-images are never zero, being always higher than the contrasts of the partial images (Fig. 1b).

Our numerical calculations show, that the contrasts of the negative and positive $\varphi-$images (Fig. 1c) are also not zero, being slightly lower than the contrast of the $t$-image, but considerably larger than the contrast $k^s$, Eq.(21) of the image, obtained by the simple fusion, Eq.(22).

Another important advantage of phase $t$-images is that, according to Eqs. (18), (19) their contrasts do not depend on the weight parameters of the image brightness, as it is the case of simple addition, Eq. (21). For the $t$-images, the contrast is determined only by the contrasts of the partial images.

It should be noted that the phase algorithms (13)-(16) of complex image fusion involve the operation of arithmetic division, which can lead to the singularity if the denominator is zero. Such a situation is quite plausible, taking into account that the input images are often obtained in conditions of poor visibility, and therefore some areas of the image might be completely extinguished such that the brightness values in the corresponding pixels are zero. In such a case, in a given point, there will be uncertainty in the brightness either for $t_{pos}$ or $t_{neg}$ value, if zero brightness is present in the partial $\upsilon-$ or $u-$image, or in both, if the brightness zero values occur in both $\upsilon-$ and $u-$images simultaneously. For $\varphi$-images, the problem of zero values in the input partial images does not show up at the analytical consideration, because $\tan z/0 = \tan\infty = \pi/2$, for any real number $z \neq 0$. However at numerical calculations, a computer program reports an error message, even in this case. In the case of zero values in the tables $\upsilon-$ or $u-$ images, we add a small parameter $0 < \varepsilon << 1$ in the denominator in expressions (13)-(16):

$$t^\varepsilon_{neg} = \tan\varphi^\varepsilon_{neg} = \frac{\upsilon}{u+\varepsilon}, \qquad (22)$$

$$t^\varepsilon_{pos} = \tan\varphi^\varepsilon_{pos} = \frac{u}{\upsilon+\varepsilon}, \qquad (23)$$

$$\varphi^\varepsilon_{neg} = \arctan\left(\frac{\upsilon}{u+\varepsilon}\right), \qquad (24)$$

$$\varphi^\varepsilon_{pos} = \arctan\left(\frac{u}{\upsilon+\varepsilon}\right). \qquad (25)$$

It turns out that the parameter $\varepsilon$ allows one to adjust the weight of the image, which is in the denominator in expressions (22)-(25). In fact, at $\upsilon << \varepsilon$ the contribution of the image, which is in the denominator can be neglected. This feature will be discussed in detail in the next section.

### 3. Examples of image fusion

Input partial images, obtained using the visual camera Nikon D3300 ($u-$image, Fig. 2a) and the thermal imaging sight ARCHER TSA-9/75-640 ($\upsilon-$image, Fig. 2b), were fused by the simple fusion ($s-$image, $\psi^s = u+\upsilon$, Fig. 3) and complex function algorithms (Fig.4).

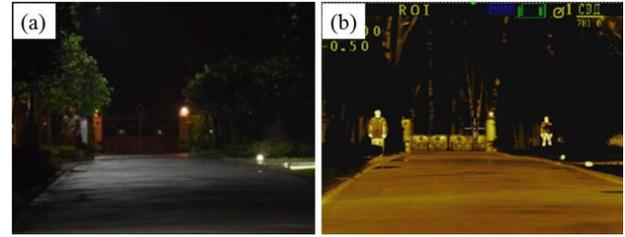

*Fig. 2 (Color online).* **Input partial images: (a) $u-$image from the visible channel and (b) $\upsilon-$ image from the IR channel**

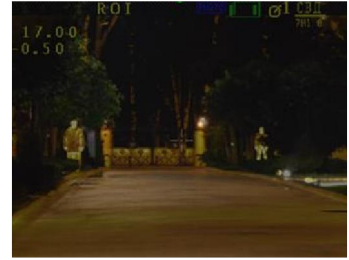

*Fig. 3 (Color online).* **The $s-$ image, obtained by the by the algorithm of simple fusion $\psi^s = u+\upsilon$ from the partial input images shown in Fig. 2a,b**

As it was noted in the previous section, the complex function approach provides the possibility to synthesize a complex image by several different algorithms. In Fig 4 two groups of images are shown, namely: positive $t_{pos} = u/\upsilon-$image (Fig. 4a), negative $t_{neg} = \upsilon/u-$image (Fig. 4b) and the positive image, obtained by standard inversion from a negative $t_{neg} = \upsilon/u-$image (Fig. 4c) as well as three $\varphi-$images: positive $\varphi_{pos} = \arctan(u/\upsilon)-$image (Fig.4d), negative $\varphi_{neg} = \arctan(\upsilon/u)-$image (Fig. 4e) and positive image obtained by standard inversion from the negative $\varphi_{neg} = \arctan(\upsilon/u)-$image (Fig. 4f).

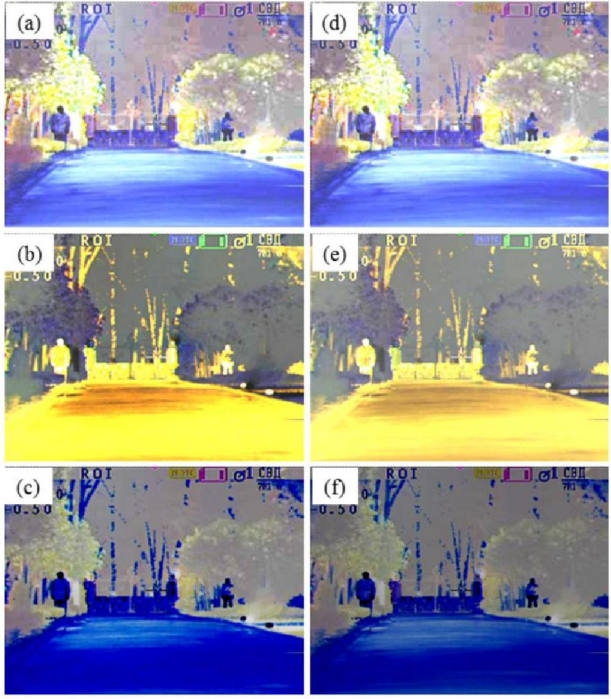

*Fig. 4 (Color online).* **Fused images by the complex function algorithm: (a) positive** $t_{pos} = u/\upsilon -$ **image, (b) negative** $t_{neg} = \upsilon/u -$ **image, (c) inverted (positive)** $t_{invpos}$ **obtained from negative** $t_{neg} = \upsilon/u -$ **image, (d) positive** $\varphi_{pos} = \arctan(u/\upsilon) -$ **image, (e) negative** $\varphi_{neg} = \arctan(\upsilon/u) -$ **image, (f) inverted (positive)** $\varphi_{invpos} -$ **image obtained from a negative** $\varphi_{neg} = \arctan(\upsilon/u) -$ **image**

It is important to note that these images were not subjected to any other processing than those provided by the corresponding Eqs. (13)-(16). Using the negative $t-$ (Fig. 4b) and $\varphi-$images (Fig. 4e), we have performed the procedure of their inversion from negative to positive using the Photoshop software. Namely, the positive images $t_{invpos} -$ (Fig. 4c) and $\varphi_{invpos} -$ images (Fig. 4f), which are shown in Fig. 4c and 4f, were obtained from the negative $t_{neg} -$ (Fig. 4b) and $\varphi_{neg} -$ images (Fig. 4e) by their inversion. It is seen that the inverted positive $t_{invpos} -$ (Fig. 4c) and $\varphi_{invpos} -$ images (Fig. 4f) obtained using the Photoshop inversion algorithms are different from the positive $t_{pos} -$ (Fig. 4a) and $\varphi_{pos} -$ (Fig. 4d) images obtained by the algorithms (14) and (16), respectively.

The inversion operation can also be applied to the positive $t_{pos} -$ (Fig. 4a) and $\varphi_{pos} -$ (Fig. 4d) images. A comparison of features of phase $t-$ and $\varphi-$images to their inverted counterparts also deserves further studies. The inverted images can also be used as fused images for further analysis. The advantages and drawbacks of the corresponding algorithms should be discussed regarding the application needs. This will be the subject of our future work.

In addition to the phase $t-$ and $\varphi-$images, one can also build the amplitude $a-$image, Eq. **Error! Reference source not found.**, i.e. by the algorithm

$$|\psi| = \sqrt{u^2 + \upsilon^2} \qquad (26)$$

and a group of combined positive and negative amplitude-phase images according to the algorithms (9), (10); they will be explored in our further studies.

Therefore, the fusion of images from visible and infrared channels by the algorithm of the complex function is a platform for the synthesis of the fused images by, at least, more than ten different algorithms. A comparison of the features of the images obtained by these algorithms requires special study and will be the subject of our next papers. In this paper, we have focused on the analysis of the phase images (Fig. 4) and their comparison to the input partial images (Fig. 2) and to the *s*-image obtained by the simple addition (Fig. 3).

Visual comparison of the input partial (Fig. 2) images to the $s-$image obtained by the simple addition (Fig. 3) is in agreement with the conclusions of the previous section according to which the simple fusion increases the brightness, but the contrast of the fused image is not higher than that of one of the partial images. The advantage of a fused image is its higher informativeness. The set $\Psi$ of elements of the fused $s-$image is the union of the sets $u$ and $\upsilon$ of the elements of the input images, namely $\Psi = u \cup \upsilon$. Indeed, in the $s-$image, one finds two persons, while the visible image (Fig. 2a) shows only one person (on left). The $\upsilon-$image from the thermal camera shows both figures, but the background details are indistinguishable, whereas details such as crowns and tree trunks (and others) are visible in the visible as well as in the fused images. Therefore, the application of the image fusion by the simple addition algorithm can be justified by the higher informativeness and higher brightness of the fused image. However, the reduction of the contrast of the fused image (Fig. 3) in comparison with that of one of the partial images (Fig. 2) is a significant drawback of the simple image fusion algorithm. This drawback is not eliminated also by the employment of the weighted fusion algorithm [2].

One can see in Fig. 4 that the contrast of the fused positive and negative phase $t-$ and $\varphi-$ images is much higher than the contrasts of the partial input $u-$ and $\upsilon-$ images. Due to the significant increase in the contrast of the positive $t-$ (Fig. 4 a) and $\varphi-$images (Fig. 4 d), one observes the non-uniformity of the background in the form of rectangles in the upper center of the images. Visual inspection of the partial $u-$image by increasing its brightness in Photoshop clearly indicates that these inhomogeneities (rectangles) are present in the input $u-$ image and are not artifacts of the image fusion. The appearance of these rectangles is the result of camera noise on the background of a weak incoming signal. Due to the increased contrast of $t$- and $\varphi-$ images, this noise becomes clearly visible on the fused image. Below we show that the visibility of the camera noise can be significantly lowered by the introduction of the so-called $\varepsilon-$ parameter in the denominator in expressions (13)-(16), which transform into Eqs. (24)-(27).

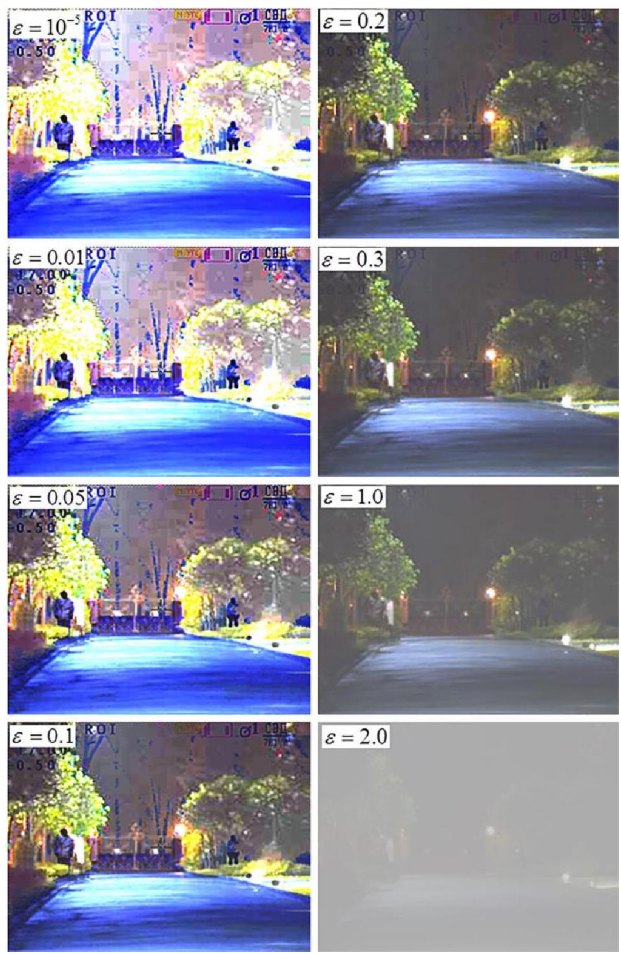

*Fig. 5 (Color online).* **Positive** $t_{pos} = u/v -$ **images, which is shown in Fig. 4a, at different values of the parameter** $\varepsilon$

Visual comparison of the $t-$ (Fig. 4 a, b) and $\varphi-$ images (Fig. 4 d, e) to the partial visual (Fig. 2a) and IR (Fig. 2b) as well as to the $s$-mage (Fig.3), confirms the conclusion, theoretically drawn in the previous section, that the contrast of the $t-$ images is significantly higher than that of the partial input images and the $s-$ image. To set the estimation of the quality of the fused images on a quantitative basis we have calculated the histograms (Fig. 6) for the partial images (Fig. 2), the $s$-image (Fig. 3) and the positive phase $t$-image (Fig. 4) after their conversion to the gray-scale format. The conclusion about the higher contrast of the phase images in comparison with that of the input partial images and of the $s$-image is supported by the histograms shown in Fig. 6a. Indeed, the widest histogram with the specific peak-like features for the $t$-image (Fig. 4) unambiguously points to its higher contrast in comparison with other images characterized in Fig. 6a.

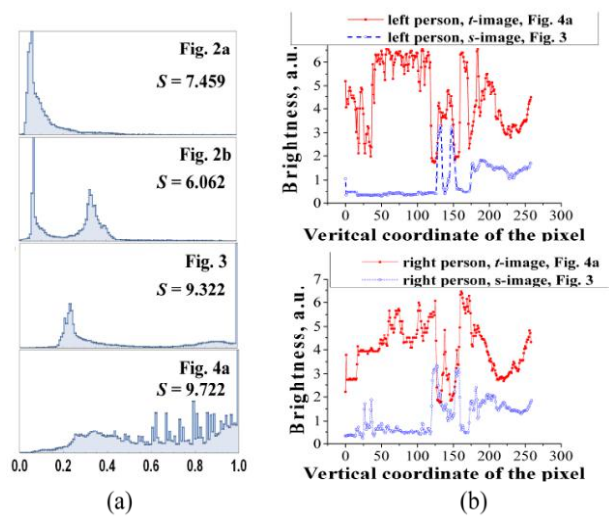

*Fig. 6.* **(a) Histograms for the partial visual (Fig.2a), IR (Fig.2b) and fused $s$- (Fig.3), phase $t$-images; (b) brightness profiles, measured along vertical lines directed through the heads of the left and right persons in Figs. 3 and 4a**

It is also worth to note that the fusion of the partial input images by the phase algorithms, Eqs. (15)-(18) and (24)-(27) considerably increases the brightness of the fused phase images in comparison with those of the input partial images and the $s$-image. The visual evidence for this conclusion is quantitatively supported by the comparison of the brightness profiles (Fig. 6b), measured along the vertical lines directed, for example, through the heads of the left and right persons in Figs. 3 and 4a. In many image processing programs (Mathematica [12] is an example), the maximal brightness value is normalized to 1 and thus the brightness values for the partial images appear to be much smaller than 1. Therefore, it is clear from Eqs. (15)-(18) and (24)-(27) that the high brightness of the phase images is due to the operation of the arithmetic division of the brightness of one partial image by that of the second one.

To estimate the amount of information in the image we have calculated the entropy $S$ of the fused images and compared these data to such data obtained for the partial input images. The notion of entropy is considered as a measure of the amount of information, from the time, when it was first introduced by Claude Shannon [13] and nowadays it is widely used for EQFI; higher entropy indicates a higher amount of information [7]. Namely, the entropy was found to be 7.459 for the input visual image (Fig. 2a), 6.062 for the input IR image (Fig. 2b), 9.322 for the *s*-image (Fig. 3), and 9.722 for the phase *t*-image (Fig. 4a). The latter indicates that the fused *s*-image contains a higher amount of information in comparison with that of the partial images, whereas the highest value of the entropy among the measured entropy values is obtained for the phase *t*-image (Fig. 4a).

To compare the quality of the proposed fusion via the algorithm of complex function to those obtained in the literature using different fusion methods, we have fused the partial images of the popular image "NATO camp" [7,14]. The histograms calculated for the *s*- and *t*-images and the corresponding entropy values $S$ are shown in Fig. 7.

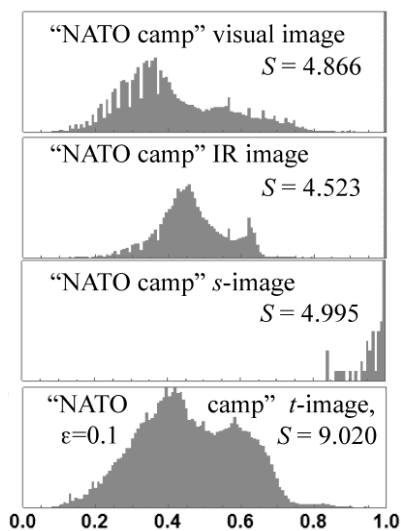

*Fig. 7.* **Histograms of the "NATO camp" partial and fused images. The value of $S$ indicates the entropy of the image.**

The histograms of the input partial images are similar, being in a form a compact band with a well-expressed sharp apex, which indicates that their brightness values are closely distributed around the mean brightness value, which in turn manifest the relatively low contrast. The histogram of the *s*-image, obtained by the simple addition, indicates that the brightness of the fused *s*-image saturates and its contrast significantly worsens in comparison with the input images. The histogram of the phase *t*-image is much wider in comparison with those of the partial images and the *s*-image, which shows that the contrast of the *t*-image is much higher. Better quality of the phase *t*-image is also confirmed by the corresponding entropy values measured for the images, shown in Fig. 7. The entropy value of the *t*-image is almost twice in comparison with those of the partial images and the *s*-image, which again indicates better quality of the *t*-image. According to the data from [7] the entropy measured for the images "NATO camp", fused by eighteen different methods falls in the range between 6.3 and 7.4. The value $S=9.020$ measured for our *t*-image appears to be considerably higher. Theoretical reasons for the higher entropy of the phase images in comparison with other methods reported in the literature as well as the behavior of other image quality indices for the phase images deserve special studies and are in progress.

It is important to note that the contrast of the $\varphi-$ images is also always higher than the contrast of the input partial images. In addition, the enhancement of the contrast is accompanied by the clearly visible increase in the brightness of the fused image.

From Fig. 4 one finds that there is a well-visible difference between the colors of the original partial images and those obtained using the *t*- and $\varphi$- lgorithms: the positive *t*- (Fig. 4a) and $\varphi$ -images (Fig. 4d) are bluish, while the negative *t*- (Fig. 4a) and $\varphi$ -images (Fig. 4d) are yellowish. In some sense these colors can be considered as pseudo colors, since they result from the mathematical operation (division and ArcTan) applied to the data from the partial images. The color appearance can be changed by the application of different possibilities for the lightening options using the condition "ColorSpace" in Mathematica. However, it is worth to note that, Fig. 5 demonstrates that the color appearance can be modified towards the natural colors of the visible image, preserving the details of both partial images by the variation of the parameter $\varepsilon$, see Eqs . (24)-(27) and Fig. 5.

Also it should be noted, that during the synthesis of the fused images presented in Fig. 4 we have faced a problem of singular values of ratio $u/\upsilon$ or $\upsilon/u$ when respectively either $\upsilon=0$ or $u=0$ in the input partial images. Image processing program used to synthesize the fused images by the algorithms (13)-(16), produces an error message due to the uncertainty at the division by zero. We eliminate this problem in the manner indicated at the end of the previous section by adding a numerical parameter $\varepsilon$ to the denominator. Namely, instead of Eqs. (13)-(16) we employ Eqs. (22)-(25).

It is important to note that there are no principal mathematical constraints on the value of the parameter $\varepsilon$. The increase of the $\varepsilon$ -value reduces the contribution of the brightness of the input partial image $\upsilon$ or $u$ standing in the denominator of the ratio $u/\upsilon$ or $\upsilon/u$, respectively. In fact, the parameter $\varepsilon$ plays the role of a weight parameter for the brightness of the image, which is in the denominator. On increasing, the parameter $\varepsilon$ reduces the brightness of the fused image.

In Fig. 5 we present the fused images for different values of $\varepsilon$ from $\varepsilon=10^{-5}$ to $\varepsilon=2$. Small values of $\varepsilon$, for example $\varepsilon=10^{-5}$ and 0.01, indeed, allow one to escape the division by zero. In addition, such small $\varepsilon$ values do not affect the contrast considerably. It can be seen from Fig. 5, that at small values $\varepsilon=10^{-5}$ and 0.01 the corresponding images are very little different from each other. Growth of $\varepsilon$ is accompanied by the approaching of the fused image to the input partial image, which stands in the nominator in Eqs. (22)-(25), let it be the visible $u$ – image, but it retains the specific details (for example, the person on the left), which are invisible in the $u$ – image (Fig. 2a), but are seen (Fig.2b) in the $\upsilon$ – image (standing in the denominator). In other words, by controlling the parameter $\varepsilon$, one can transform the fused image closer to the natural appearance of the photos in the visible light, without losing the information that appeared on it due to the image fusion. Compare, for example, the original image from the visual camera, shown in Fig. 2a to the fused $t$-image with $\varepsilon = 0.2$ in Fig.5. It is clearly seen that the fused $t$-image with $\varepsilon = 0.2$ in Fig. 5 is of much higher quality than the original image shown in Fig. 2a: the $t$-image is of higher contrast and contains details (figure of a second man on the right, for example, but not only), which are not seen in the original image. In this sense, the fusion by the algorithm of the complex function is akin to the algorithm of weighted fusion but with significantly higher contrast, brightness and amount of the information.

It is important also to note the visibility of the noise in the image from visible camera in the form of bright and dark rectangles can be reduced increasing the value of the parameter $\varepsilon$ (compare the image with $\varepsilon = 10^{-5}$ to that, for example, with $\varepsilon = 0.2$ in Fig. 5).

Finally, it should be noted that the proposed method is designed for the purpose of the target recognition and implies minimization of the subjective influence of an operator. Subjective methods of EQFI are of great importance for civil applications, where the art-component prevails. For the scientific (including medical) and military applications the application of the subjective influence of an operator and methods should be minimized. The proposed method is designed for the scientific and military needs and for this reason we do not apply the subjective methods of EQFI to its evaluation.

## Conclusions

We propose an algorithm for the fusion of partial images collected from the visual and infrared cameras in the form of a complex function such that one of the partial images is the real part while the other one is the imaginary part of the complex function. Since there is no restriction on the choice which of the two images will be the real and which one will be the imaginary part, the fused image can be built either as $\psi_{neg} = u + i\upsilon$ or $\psi_{pos} = \upsilon + iu$, where $u$ and $\upsilon$ are the images from the visual and infrared channels, respectively. Rewriting the complex function to the exponential (or trigonometric) form:

$$\psi_{neg,pos} = |\psi|(\sin\varphi_{neg,pos} + i\cos\varphi_{neg,pos}) = |\psi|e^{i\varphi_{neg,pos}}$$

provides two possibilities to synthesize the fused image as either the amplitude $|\psi| = \sqrt{u^2 + \upsilon^2}$ or the phase images, which can be built either as the two $t$ – images: $\tan\varphi_{neg} = \upsilon/u$, $\tan\varphi_{pos} = u/\upsilon$, or two $\varphi$ – images: $\varphi_{neg} = \arctan(\upsilon/u)$ and $\varphi_{pos} = \arctan(u/\upsilon)$. Unlike the phase images, the appearance of the amplitude image does not depend on the choice of the real and imaginary parts of the complex function algorithm. In the above notations the subscripts "*neg*" and "*pos*" correspond, respectively to the negative and positive format of the image in their conventional senses.

One of the advantages of the proposed complex function algorithm is that the multiple images from the two channels are sorted and fused separately, let say by the weighted addition algorithm and then these two fused images from the two channels are used to form either the amplitude or phase images.

We show that due to the fact that the local contrasts of the partial visual and infrared images are of the opposite signs, the local contrast of the fused phase $t$ – image is the sum of the absolute values of the local contrasts of the partial images, whereas the local contrast of the image fused by the algorithm of simple adding is the difference between the local contrasts of the corresponding partial images. Thus, the local contrast of the fused phase image is higher than those of the partial images as well as of the image fused by the simple or weighted addition algorithm. Due to the opposite signs of the contrasts of the visible and infrared images the contrast of the image fused by the algorithm of simple addition can vanish, whereas the algorithm of complex function does not have such a drawback. Theoretical conclusions are supported by the computer modeling as well as by the qualitative (visual) and quantitative characterization of the examples of the fusion of real visual and infrared images using the proposed algorithm of complex function and the algorithm of simple addition.

We believe that the enhancement of the local contrast in the fused phase images is also promising for application to the problem of object recognition.

## References

*1. Castanedo F. A Review of Data Fusion Techniques / F. Castanedo // The ScientificWorld Journal. 2013. 19p. Article ID: 70450, DOI: http://dx.doi.org/10.1155/2013/704504*

*2. Khaustov Ya.Ye. Image fusion for a target sightseeing system of armored vehicles / Ya.Ye. Khaustov, D.Ye. Khaustov, E. Lychkovskyy, Ye. Ryzhov, Yu.A. Nastishin // Development and modernization military equipment. Military Technical Collection – Lviv, 2019. no. 21 P. 2837. DOI: https://doi.org/10.33577/2312-4458.21.2019.28-37* .

3. Kong W. Adaptive fusion method of visible light and infrared images based on non-subsampled shearlet transform and fast non-negative matrix factorization / W. Kong, Y. Lei, H. Zhao // Infrared Physics & Technology. 2014. vol. 67. P. 161–172. DOI: http://dx.doi.org/10.1016/j.infrared.2014.07.019

4. Ma J. Infrared and visible image fusion via gradient transfer and total variation minimization / J. Ma, Ch. Chen, Ch. Li, J. Huang // Information Fusion. 2016. vol. 31. P. 100–109. DOI: http://dx.doi.org/10.1016/j.inffus.2016.02.001

5. Ma J. FusionGAN: A generative adversarial network for infrared and visible image fusion / J. Ma, W. Yu, P. Liang, Ch. Li, J. Jiang // Information Fusion. 2019. vol. 48. P. 11-26. DOI: https://doi.org/10.1016/j.inffus.2018.09.004

6. Ma J. Infrared and visible image fusion via detail preserving adversarial learning / J. Ma, P. Liang, W. Yu, Ch. Chen, X. Guo, J. Wu, J. Jiang // Information Fusion. 2020. vol. 54. P. 85-98. DOI: https://doi.org/10.1016/j.inffus.2019.07.005

7. Ma J. Infrared and visible image fusion methods and applications: A survey / J. Ma, Y. Ma, C. Li // Information Fusion. 2019. vol. 45. P. 153-178. DOI: https://doi.org/10.1016/j.inffus.2018.02.004

8. Zitova B. Image Registration Methods: A survey / B. Zitova, J. Flusser // Image and Vision Computing. 2003. vol. 21. P. 977-1000. DOI: https://doi.org/10.1016/S0262-8856(03)00137-9

9. Gonzalez R.C. Digital Image Processing / C. Gonzalez, R. E. Woods and S. L. Eddins. Pearson Education, 2003. 976p.

10. Jolliffe I.T., Principal Component Analysis, 2nd ed. / I.T. Jolliffe. Springer-Verlag, Berlin, Germany, 2002 518p. ISBN 0-387-95442-2.

11. Markushevich A.I. Theory of functions of a complex variable, 2nd ed. (3 vol. set) / A. I. Markushevich. American Mathematical Society, 2017. 1138p.

12. Wolfram S. The Mathematica Book: 5th ed. / S. Wolfram. New York : Wolfram Media, 2005. 1486p.

13. Shannon C.E. A mathematical theory of communication / C.E. Shannon // The Bell System Technical Journal. 1948. vol. 27. P. 379–423. DOI: https://doi.org/10.1002/j.1538-7305.1948.tb01338.x

14. Huang G. Visual and infrared dual-band false color image fusion method motivated by Land's experiment / G. Huang, G. Ni, B. Zhang // Optical Engineering. 2007. vol. 46(2). DOI: https://doi.org/10.1117/1.2709851

**Комплексування зображень з видимого та тепловізійного каналів у формі комплексної функції**

Я.Є. Хаустов, Д.Є. Хаустов, Є.В. Рижов, Е.І. Личковський, Ю.А. Настишин

*У даній роботі ми пропонуємо алгоритм комплексування парціальних зображень, отриманих з видимої та інфрачервоної камер у формі комплексної функції, що є простим у використанні та невимогливим до потужностей сучасної електронно-обчислювальної техніки. Алгоритм комплексування зображень через комплексну функцію є узагальненням алгоритму додавання зображень так само, як додавання комплексних чисел є узагальненням додавання дійсних чисел. Однією з переваг запропонованого алгоритму комплексної функції є те, що кілька зображень з двох каналів сортуються та комплексуються окремо, скажімо, за алгоритмом вагового додавання, а потім ці два комплексованих зображень з двох каналів використовуються для формування амплітудного чи фазового зображення. Комплексна форма комплексованого зображення відкриває можливість сформувати комплексоване зображення як амплітудне зображення або як фазове зображення, яке, у свою чергу, може бути в декількох формах.*

*Ми показуємо, що через те, що локальні контрасти часткових візуальних та інфрачервоних зображень мають протилежні ознаки, локальний контраст комплексованого фазового зображення є сумою абсолютних значень локальних контрастів парціальних зображень, тоді як локальний контраст комплексованого зображення, отриманого за алгоритмом простого додавання, − це різниця між локальними контрастами відповідних парціальних зображень. Таким чином, локальний контраст комплексованого фазового зображення вищий, ніж у парціальних зображень, а також у порівнянні із зображеннями, комплексованими за допомогою алгоритмів простого або вагового додавання. Через протилежні знаки контрастів видимих та інфрачервоних зображень контраст зображення, комплексованого за алгоритмом простого додавання, може зануилитись, тоді як алгоритм комплексної функції не має такого недоліку. Теоретичні висновки підкріплені комп'ютерним моделюванням в середовищі Mathematica, а також на прикладах комплексування реальних видимих та інфрачервоних зображень за допомогою запропонованого алгоритму комплексної функції та алгоритму простого додавання.*

**Ключові слова:** *цифрова обробка зображень, комплексування зображень, тепловізійне бачення, оцінка якості зображення.*

**Комплексирование изображений видимого и тепловизионного каналов в форме комплексной функции**

Я.Е. Хаустов, Д.Е. Хаустов, Е.В. Рыжов, Э.И. Лычковский, Ю.А. Настишин

*В представленной работе мы предлагаем алгоритм комплексирования парциальных изображений, полученных видимой и инфракрасной камерами, в виде комплексной функции, которая является простой в использовании и не требовательной к мощностям современной электронно-вычислительной техники. Алгоритм комплексирования изображений через комплексную функцию является обобщением алгоритма слияния изображений так же, как сложение комплексных чисел является обобщением сложения действительных чисел. Одним из преимуществ предложенного алгоритма комплексной функции есть то, что несколько изображений из двух каналов сортируются и комплексируются отдельно, скажем, по алгоритму весового сложения, а потом эти два комплексированных изображения из двух каналов используются для формирования амплитудного или фазового изображения. Комплексная форма комплексированного изображения открывает возможность сформировать комплексное изображение как амплитудное изображение или как фазовое изображение, которое, в свою очередь, может быть в нескольких формах.*

*Мы показываем, что из-за того, что локальные контрасты парциальных визуальных и инфракрасных изображений имеют противоположные знаки, локальный контраст комплексированного фазового изображения является суммой абсолютных значений локальных контрастов парциальных изображений, тогда как локальный контраст комплексированного изображения, полученного по алгоритму простого сложения, − это разница между локальными контрастами соответствующих парциальных изображений. Таким образом, локальный контраст комплексированного фазового изображения выше, чем у парциальных изображений, а также в сравнении с изображениями, комплексированными с помощью алгоритмов простого или весового сложения. Из-за противоположных знаков контрастов видимых и инфракрасных изображений контраст изображения, комплексированного алгоритмом простого сложения, может занулится, тогда как алгоритм комплексной функции не имеет такого недостатка. Теоретические выводы подкреплены компьютерным моделированием в среде Mathematica, а также на примерах комплексирования реальных видимых и инфракрасных изображений с помощью предложенного алгоритма комплексной функции и алгоритма простого сложения.*

***Ключевые слова:*** *цифровая обработка изображений, комплексирование изображений, тепловизионное видение, оценка качества изображения.*